\documentclass[letterpaper, 10 pt, conference]{ieeeconf}
\IEEEoverridecommandlockouts

\def\BibTeX{{\rm B\kern-.05em{\sc i\kern-.025em b}\kern-.08em
    T\kern-.1667em\lower.7ex\hbox{E}\kern-.125emX}}
\usepackage{epsfig} 
\usepackage{url}
\usepackage{hyperref}
\usepackage{cite}
\usepackage{amsmath,amssymb,amsfonts,amsthm}
\usepackage{algorithmic}
\usepackage{graphics}
\usepackage{textcomp}
\usepackage{xcolor}
\usepackage{enumitem}
\usepackage{soul}
\usepackage{subcaption}
\usepackage{graphicx}

\newcommand{\rs}[1]{R_{s_{#1}}}
\newcommand{\rc}[1]{R_{c_{#1}}}
\newcommand{\xs}[1]{x_{s_{#1}}}
\newcommand{\Ss}{\mathcal{S}}
\newcommand{\F}{\mathcal{F}}
\newcommand{\N}{\mathcal{N}}

 \renewcommand{\QED}{\QEDclosed}

\newtheorem{thm}{Theorem}
\newtheorem{df}{Definition}

\begin{document}

\title{A Distributed Strategy to Maximize Coverage in a Heterogeneous Sensor Network in the Presence of Obstacles
\thanks{
This work has been supported by the Natural Sciences and
Engineering Research Council of Canada (NSERC) under Grant
RGPIN-262127-17.}
}

\author{Hesam Mosalli and Amir G. Aghdam
\thanks{H. Mosalli and A. G. Aghdam are with the Department of Electrical and Computer Engineering, Concordia University, Montréal, QC, Canada. Email:
        {\tt\small hesam.mosalli@mail.concordia.ca}, {\tt\small amir.aghdam@concordia.ca}}%
}


\maketitle

\begin{abstract}
In this paper, an efficient deployment strategy is proposed for a network of mobile and static sensors with nonidentical sensing and communication radii.  The multiplicatively weighted Voronoi (MW-Voronoi) diagram is used to partition the field and assign the underlying coverage task to each mobile sensor. A gradient-based method is applied to find the best candidate point based on the detected coverage holes and the coverage priority considering the relative distance of the mobile sensor from the static ones and the obstacles in the field. The sensors move to a new position if such a relocation increases their local coverage. The efficiency of the proposed strategy in different scenarios is demonstrated by simulations.
\end{abstract}


\section{Introduction}
Wireless Sensor Networks (WSNs) have emerged as a promising technology for various applications such as environment monitoring, healthcare, and surveillance~\cite{lin,pottie, tubaishat,Kim2021}.
However, a critical challenge in WSNs is ensuring that the sensors adequately cover the area of interest to guarantee reliable data collection. The problem is particularly important in scenarios involving heterogeneous nodes and fixed obstacles in the field, e.g., when the sensing field is a harsh outdoor environment\cite{Chatzigiannakis}. Various deployment strategies have been introduced in the literature to address this challenge.

Distributed deployment strategies in a WSN aim to move every sensor in a field to increase the covered area with minimal information exchange with other sensors.
They
often use a Voronoi-based approach to partition the sensing field into regions and assign a node (sensor) to each
(for the mathematical description of the Voronoi diagram, see \cite{okabe}).
Virtual force-based algorithms are proposed in \cite{mahboubi2017,wang2006, zou2003} to move the mobile sensors to enhance the covered area in a WSN. These algorithms use a combination of attraction and repulsion forces to relocate sensors.

Gradient-based approaches are another class of coverage maximization strategies in mobile WSNs. These approaches use the gradient of the sensing field to determine each sensor’s optimal moving direction and size. These algorithms may also consider environmental obstacles and constraints, such as energy consumption or the communication ranges of the sensors
making
such methods suitable for generalizing the problem statement and applying the necessary modifications in the strategy. A gradient descent algorithm is proposed in \cite{cortez2004} for a class of utility functions encoding the optimal coverage and sensing policies. Moreover, \cite{habibi2017} utilizes a distributed nonlinear optimization approach iteratively to
increase the local coverage
of each sensor
as much as possible.

Many real-world WSNs are heterogeneous, i.e., sensors have different
characteristics. Heterogeneity poses additional challenges for maximizing the covered area in a mobile WSN. In some studies, the multiplicatively weighted Voronoi (MW-Voronoi) diagram \cite{deza} is employed to partition the field according to the sensors' sensing radii to maximize coverage~\cite{mahboubi2014}. There are limited studies considering a network of both mobile and static sensors,
and most of them
formulate the static sensors as sink nodes in the network to maximize coverage in two steps: allocation of the static sensors and path planning of the mobile sensors~\cite{guo}.
On the other hand, the presence of obstacles in the field can negatively impact the functionality of the WSN. An efficient obstacle detection scheme is proposed in  \cite{wang2013}, which models the obstacles as coverage holes. The centric MW-Voronoi configuration introduced in \cite{mahboubi2019} guarantees the convergence of the mobile sensors in a heterogeneous network to the optimal locations in the presence of obstacles and limited communication ranges.

In this paper, a gradient-based distributed strategy is introduced to solve the weighted coverage optimization problem. Unlike the previous studies, the proposed approach can address four problems simultaneously, i.e., network heterogeneity, the prioritized sensing field, the existence of obstacles in the field, and the presence of both mobile and static sensors. In an iterative procedure, each sensor first uses the information received from its neighbors to construct its Voronoi region. Then, it determines its optimal location by solving the local coverage maximization problem.

This paper is organized as follows. The problem formulation
is
presented in Section~\ref{sec2}, along with some preliminary concepts. In Section~\ref{sec3}, coverage maximization is formulated as a nonlinear optimization problem, and a distributed algorithm is provided to solve it. An alternative approach, namely the modified max-area strategy, is provided in Section~\ref{sec4}. The performance of the strategy is demonstrated by simulations in Section~\ref{secex}. Finally, Section~\ref{secconc} concludes the paper by summarizing the results.

\section{Preliminaries and problem statement}
\label{sec2}
Consider a 2D sensing field $\F$, which is to be covered by a set of $n$ sensors
$\Ss=\left\{
S_i(\xs{i}, \rs{i}, \rc{i}) \vert i\in \mathbb{N}_n\right\}$, $\mathbb{N}_n :=\{1, 2, \dots, n\}$, where
$\xs{i}$, $\rs{i}$, and $\rc{i}$
are, respectively, the position, sensing radius, and communication radius of sensor $S_i$.
Each sensor
$S_i$
is assumed to have a disk-shaped sensing range with the radius $\rs{i}$, capable of broadcasting its location and sensing
radius to other sensors in its communication radius.
Assume also that sensors
$S_1, S_2, \dots, S_m$ are mobile and the remaining $n-m$ are static, i.e., their positions are fixed. Also, the sensing and communication radii of different sensors are not necessarily the same.
A priority function
$\varphi(q): \F \to \mathbb{R}^+$, where $\mathbb{R}^+$ is the set of all non-negative real numbers, describes the relative importance of
point $q$
inside the 2D field to be covered. A point with a higher value of the priority function is more important to cover compared to a point with a lower value.

The region of interest (ROI) may include fixed obstacles with arbitrary shapes that block the sensing range of the sensors. It is assumed that (i) the effect of obstacles on wireless communication between sensor nodes is negligible and can be compensated via multi-path signal propagation~\cite{Chatzigiannakis}, and (ii) each sensor is capable of detecting the exact shape of any obstacle within its communication range. To analyze the effect of obstacles on the coverage performance of the sensor network, the \emph{visible region} of a sensor is defined below.

\begin{df}
Consider a sensing field with some obstacles. The visible region of a sensor located at point $x$ is denoted by $\Phi(x)$ and includes all the points in $\F$ from which there exists an unobstructed line of sight to $x$. 
\end{df}

\begin{df}
Since it is assumed that obstacles block the sensor's line of sight, the sensing range of the sensor $S\left(x, \rs{}, \rc{}\right)$ is defined as:
\begin{equation}
    D(x) = \left\{q\in\Phi(x)\vert d(q,x)	\leq \rs{}\right\},
\end{equation}
where $d(q,x)$ is the Euclidean distance between points $q$ and $x$.
\end{df}

It is desired to find a set of locations for the sensors resulting in the maximum coverage over the ROI. To achieve this goal, a distributed strategy is proposed under which each mobile sensor uses the information it obtains from its neighbors to iteratively find a new point from which its local coverage increases. The local coverage optimization problem is solved using a Voronoi-based approach by partitioning the ROI into preferably distinct regions, each assigned to one of the $m$ mobile sensors.

In Voronoi-based approaches, the fundamental assumption is to partition the entire area of a sensing field into distinct regions such that the sensor located inside each region
is the nearest sensor to all points within that region~\cite{okabe}.
While the Voronoi diagram provides the standard partitioning in a homogeneous network consisting of sensors with identical sensing radii, the MW-Voronoi diagram presents the desired partitioning for the cases when the sensors have different sensing radii~\cite{deza}. Moreover, since the sensors have a limited communication range in general,
the notion of \emph{Connectivity-Aware Multiplicatively Weighted Voronoi} (CAMW-Voronoi) diagram is used in this paper which is slightly different from the LCMW-Voronoi diagram introduced in~\cite{mahboubi2013}.
Let the weighted distance between a point $q\in\F$ and a weighted node $S(x,w)$ be defined as:
\begin{equation}
    d_w(q,S)=\frac{d(q,S)}{w}.
\end{equation}

\begin{df}
Consider the network of sensors
$\Ss$
introduced earlier. Let the set of all neighbors of a mobile sensor $S_i$, denoted by $\N_i$, be the set of all mobile sensors whose communication ranges reach $S_i$, i.e., it can receive information from them.  Then, the connectivity-aware multiplicatively weighted Voronoi region associated with $S_i$ is defined as:
\begin{align}
    \Pi_{i}= \{& {q\in \F}\vert d_{w}(q,S_{i})<d_{w}(q,S_{j}), \forall j\in {\N_i},\nonumber\\
    & d(q,S_i)<\min\{\rc{i}, r_{min}\}\},
\end{align}
where 
$r_{min}=\min_{j\in \N_i}\left\{\left[d(S_i,S_j)-\rs{j}\right]_+\vert i\notin\N_j\right\}$, $\left([a]_+=\max\{a,0\}\right)$ and the corresponding weight of each sensor is equal to its sensing radius.
\end{df}

Note that based on the definition of the weighted distance, if a sensor cannot cover an arbitrary point inside its CAMW-Voronoi region, none of its neighbors can cover it. This is a critical point making the corresponding regions
important in developing a distributed deployment strategy for coverage optimization. Unlike the MW-Voronoi diagram, the CAMW-Voronoi diagram is not necessarily a complete partitioning of $\F$, as the regions are not always mutually distinct due to the limitation on the communication ranges of the sensors. However, it is worth mentioning that the effect of such shortcomings can be minimized in a good network configuration where the communication radii are sufficiently large. 

\emph{Problem Definition:} Given the specifications of the environment including the obstacles and the priority function, and the network configurations, it is desired to find a set of locations for the mobile sensors that achieves the maximum weighted coverage over the ROI. Here, the overall weighted coverage is defined as the surface integral of the priority function over the field $\F$. The overall weighted coverage maximization problem is formulated as follows.
\begin{equation}\max_{\left\{x_{i}\right\}_{i=1}^{m}}\quad\ \int\limits_{{\F\cap \left(\bigcup_{i=1}^{n} D\left(x_{{i}}\right)\right)}}\varphi(q)dq.
\label{eq:owc}
\end{equation}
Here, the overall coverage is computed based on the areas covered by all sensors but only mobile sensors can contribute to modifying it.

In Voronoi-based approaches, each mobile sensor is assigned the task of maximizing the local coverage w.r.t. the Voronoi region associated with itself.

\section{Distributed maximum weighted coverage problem}
\label{sec3}
Since it is not straightforward to find the globally optimal solution in a distributed strategy, it is useful to reformulate it as multiple local problems. The problem of distributed maximum weighted coverage  in a homogeneous network with no static sensors, no obstacles, and no constraint on the communication range of the sensors has been solved in~\cite{habibi2017}. In a similar way,  an iterative approach is proposed in which each sensor is able to find the best position to move to maximize its own local weighted coverage. 
Under such conditions, the overall weighted coverage of the sensor network $\Ss$ in each step could be computed as the sum of the local weighted coverage of all sensors over their associated Voronoi regions. In a similar way, the overall weighted coverage (\ref{eq:owc}) can be rewritten as:

\begin{align}
\max_{\left\{x_i,\Pi_{i}\right\}_{i=1}^{m}}\quad&\sum_{i=1}^{m}\ \ \int\limits_{\Pi'_{i}\cap D\left(x_i\right)}\varphi(q)dq\\ {\mbox{subject to}}:\quad& x_i\in\Pi_{i}\cap \Phi(\xs{i}),\quad \forall\ i\in\mathbb{N}_m,\nonumber
\end{align}
where $\Pi'_i$ is part of $\Pi_i$ that is not covered by any static sensor. It is to be noted that the above reformulation of overall weighted coverage (\ref{eq:owc}) as the sum of local weighted coverage of mobile sensors is valid if regions
$\Pi_1, \Pi_2, \dots, \Pi_{m}$
are mutually distinct (and so are the local covered areas). However, this is not the case in a CAMW-Voronoi diagram due to the possible overlap between the regions.

\begin{thm}
\label{thm: rc2rs}
Suppose that
$\left\{\Pi_1, \Pi_2, \dots, \Pi_{m}\right\}$
is the CAMW-Voronoi diagram generated by $\Ss$.  Then, the local covered areas, defined as
$\Pi_i\cap D(x_i)$, are mutually distinct for any two sensors if
$\rc{i}>2\rs{i}$ for all $i\in\mathbb{N}_m$.
\end{thm}
\begin{Proof}
The proposition will be proved for two sensors $S_1$ and $S_2$ and can then be extended to the entire network $\Ss$.  Without loss of generality, assume that $\rc{1}\leq\rc{2}$. There are three possible cases:
\begin{enumerate}
    \item If $1\in\N_2$ and $2\in\N_1$, then the covered areas of the two sensors are separated based on the definition of the CAMW-Voronoi diagram.
    \item Suppose that $1\notin\N_2$ and $2\notin\N_1$ meaning that $\rc{1}\leq\rc{2}<d(S_1,S_2)$. If the inequality $\rc{i}>2\rs{i}$ holds for both sensors, it can be deduced that $\rs{1}+\rs{2}<d(S_1,S_2)$. So, the sensing disks of the sensors do not overlap, making the covered areas distinct.
    \item Lastly,  suppose that $1\in\N_2$ and $2\notin\N_1$ (note that by assumption, we can not have $1\notin\N_2$ and $2\in\N_1$). Based on the definition of $r_{min}$, the extent of region $\Pi_1$ and consequently the covered area of $S_1$ are limited by the sensing disk of $S_2$. As a result, under the stated relation between the sensing and communication radii, the covered areas of all sensors are mutually distinct.
\end{enumerate}
\end{Proof}

Theorem~\ref{thm: rc2rs} provides 
a basis to ensure the objective function in (\ref{eq:owc}) is separable, and the overall weighted coverage maximization problem can be iteratively solved in a distributed manner as formulated in (\ref{eq:stdmin}).
Furthermore, the provided condition is almost always true in operational cases due to the inherent difference between the communication and sensing equipment and methods in WSNs \cite{mahfoudh}.

In what follows, the distributed strategy for maximizing the local weighted coverage is described for one sensor. Consider a single sensor $S$ with a sensing radius $\rs{}$ located at $\xs{}$ inside its CAMW-Voronoi region $\Pi$. The goal is to find the optimal point $x$ inside $\Pi$ for which both of the following criteria hold.
\begin{itemize}
    \item The sensor can move from its current position to $x$ on an unobstructed line.
    \item The maximum local weighted coverage of $S$ over $\Pi'$ is achieved.
\end{itemize}
Also, instead of presenting the objective in the form of a maximization problem, it is preferable to formulate it in a standard minimization problem as:
\begin{align}
\min_{x}\quad&F(x)=-\int\limits_{\Pi'\cap D\left(x\right)}\varphi(q)dq \label{eq:stdmin}\\
{\mbox{subject to}}:\quad& x\in\Pi\cap \Phi(\xs{}).\nonumber
\end{align}

\begin{figure}
    \centering
    \includegraphics[width=.8\columnwidth, draft=false]{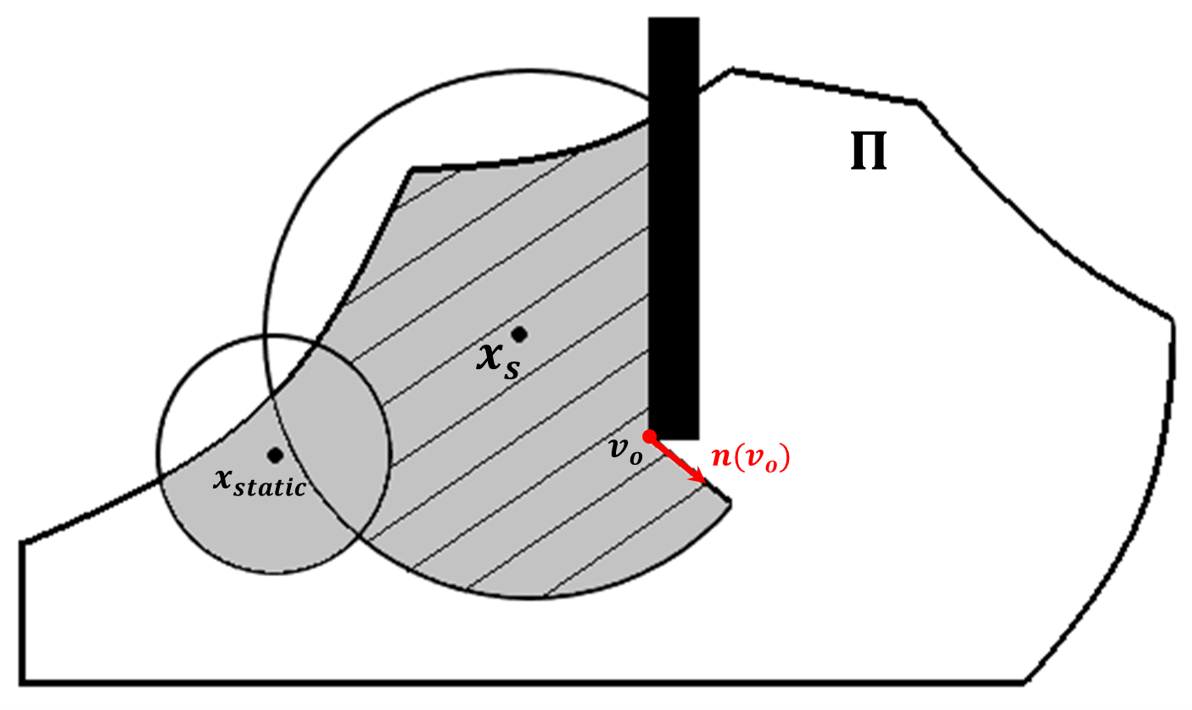}
    \caption{Geometric illustration of the local coverage maximization problem}
    \label{fig:vd}
\end{figure}

The geometric configuration of the problem is illustrated by a simple example in Fig.~\ref{fig:vd}. It is desired to find the optimal point for $S$ inside $\Pi$ such that the local weighted coverage over the hatched area is maximized, considering that the sensing disk and moving routes are blocked by the obstacle (depicted in black). Also, note that the area covered by the static sensor does not need to be considered in the coverage maximization problem.

The problem described by (\ref{eq:stdmin}) is generally a nonlinear optimization problem due to the constraints imposed by the heterogeneity of the network, the presence of static sensors, and the existence of obstacles in the field. As a result,  the gradient-based approach introduced in \cite{habibi2017} is modified to adapt it to the challenges introduced by such constraints. 

\section{Modified Max-Area Strategy}
\label{sec4}
In the modified max-area (MMA) strategy introduced in this section, the mobile sensors find their optimal locations in the field iteratively, as described in the following {algorithm}:
\begin{enumerate}[label=\roman*.]
\item Each sensor transmits information containing its location and sensing radius to the sensors within its communication range and receives the same information from its neighbors.
\item Every sensor constructs its own CAMW-Voronoi region based on the information it receives from its neighbors in the previous step and the shape of the obstacles it detects.
\item Each sensor computes its local weighted coverage over its region.
\item If a sensor detects a local coverage hole, it finds a candidate point using the gradient-based method described below.
\item The sensor computes its local weighted coverage in the candidate position w.r.t. its current Voronoi region and moves accordingly
if the new local coverage is more than the current value by a prescribed threshold $\epsilon$.
\item If at least one of the sensors moves in the previous step, the procedure repeats from step (i); otherwise, it ends.
\end{enumerate}

The above procedure is repeated iteratively until no sensor moves, i.e., the network reaches a steady state regarding the weighted coverage. The choice of $\epsilon$ in step (v) involves a trade-off between precision and convergence time. The smaller $\epsilon$ is, the closer are to their optimal positions, but the longer it takes for the algorithm to reach the termination condition, when the sensors stop. 

To solve the optimization problem defined in (\ref{eq:stdmin}), in each iteration $k$, first, the best direction to move, denoted by $p_k$, which is proportional to the gradient of the function $F(x)$, is found. Therefore, the primary step is to compute the vector $\nabla_x F(x)$. This can be performed similarly to the Max-Area strategy \cite{habibi2017}. As a more general case, suppose that the objective function to minimize is the surface integral of a given function $f$ over the region $\mu(x)$, which is the loci of every point $q\in\F$ satisfying $M$ inequalities $h_j(x,q)\leq 0, (j\in\mathbb{N}_M)$. The boundary of $\mu(x)$, denoted by $\partial\mu(x)$, consists of $M$ segments (curves or lines), where the $j$-th segment is $\partial_j\mu(x)=\left\{q\in\mu(x)\vert h_j(x,q)=0\right\}$.  In this case, the gradient of the objective function
\begin{equation}
F(x)=\int_{\mu(x)}f(x,q)dq
\end{equation}
w.r.t. $x$ can be derived using the results of \cite{uryasev} as follows:
\begin{align}   \nabla_{x}F(x)=&\int\limits_{\mu(x)}\nabla_{x}f(x,q)dq\nonumber\\ &-\sum_{j=1}^{M}\int\limits_{\partial_{j}\mu(x)}{\frac{f(x,q)}{\left\Vert\nabla_{q}h_{j}(x,q)\right\Vert}}\nabla_{x}h_{j}(x,q)dL. \label{eq:pgradient}
\end{align}

In the local weighted coverage optimization problem, the function $f$ in the above formulation is equal to the priority function $\varphi(q)$, and is independent of the position of sensor $x$, making the first term in (\ref{eq:pgradient}) equal to zero. Since the second term is a line integral over the boundaries of $\mu(x)$ (i.e., $\Pi'\cap D\left(x\right)$ in the MMA strategy), the boundary segments are first identified. The first two types of boundary segments are the boundaries of $\Pi$ and the sensing ranges of every static sensor, lying inside the sensing range of the sensor.
The third group involves parts of the obstacle edges that are facing the sensor, inside the sensing range of the sensor, and not covered by any static sensor. Without loss of generality, suppose that these three groups of boundaries include the first $p$ segments, characterized by functions $h_1, h_2, \dots, h_p$. Since these segments do not change by the movement of the sensor, thus,
\begin{equation}
    \nabla_{x}h_{j}(x,q)=0, \forall j\in\mathbb{N}_p.
\end{equation}

Another segment of the region boundary is the portion of the perimeter of the sensing disk that is inside both the visible region of the sensor and region $\Pi$ but is not covered by any static sensor. Let this segment be described by the equality $h_{p+1}(x,q)=0$. It is shown in~\cite{habibi2017} that the gradient of the objective function originating from this set can be simply computed by a discrete summation as:
\begin{align}
\nabla_{x}^{D}F(x)&=\frac{2\pi \rs{}}{N_{p+1}} \sum_{k=1\atop q^D_{k}\in\Pi'}^{N_{p+1}}\left[\begin{array}{@{}c@{}}\cos\theta_{k}\\ \sin\theta_{k}\end{array}\right]\varphi(q^D_{k}),\label{eq:gradient_disk}
\end{align}
where, $\theta_k=2(k-1)\pi/N_{p+1}$ ($k\in \mathbb{N}_{N_{p+1}}$)
and $q^D_k=x+\rs{}\left[cos\theta_k,sin\theta_k\right]^T$. The number
$N_{p+1}$ is sufficiently large to guarantee the desired precision in computing the integral in a discrete form.

The last group of segments is generated when the sensing disk is blocked by obstacles. These boundary partitions are line segments in a radial direction in the sensing disk starting from the blocking vertices of the obstacle denoted by $v_o=\left[v_{o,1},v_{o,2}\right]^T$ to the point on the perimeter of the sensing disk. This is illustrated in Fig.~\ref{fig:vd} for a simple configuration. For now, assume that there is only one such line segment and that no part of it is covered by any static sensor. Define:
\begin{equation*}
    h_{p+2}\left(x, q\right)=\left(q_2-v_{o,2}\right)\left(v_{o,1}-x_1\right)-\left(q_1-v_{o,1}\right)\left(v_{o,2}-x_2\right)
\end{equation*}
which yields:
\begin{align}
& \nabla_{x} h_{p+2}\left(x, q\right)=\left[
\begin{matrix}
-\left(q_2-v_{o,2}\right) \\
q_1-v_{o,1}
\end{matrix}\right]=\left[
\begin{matrix}
0 & -1 \\
1 & 0
\end{matrix}
\right]\left(q-v_o\right), \\
& \nabla_q h_{p+2}\left(x, q\right)=\left[
\begin{matrix}
-\left(v_{o,2}-x_{2}\right) \\
v_{o,1}-x_{1}
\end{matrix}
\right],
\end{align}
resulting in
\begin{equation}
    \left\|\nabla_q h_{p+2}\left(x, q\right)\right\|=\left\|v_o-x\right\|.
\end{equation}
For simplicity, let $\left\|v_o-x\right\|=d(x,v_o)$ be called $R_{v_o}$. By substituting these values in (\ref{eq:pgradient}), one obtains:
\begin{align}
\nabla_{x}^{v_o}F(x)=
\left[
    \begin{matrix}
    0 & -1 \\
    1 & 0 
    \end{matrix}
\right]\frac{1}{R_{v_o}}
\int\limits_{\partial_{p+2} \mu(x)} \left(q-v_o\right) \varphi(q) dL. \label{eq:gradient_obs}
\end{align}
 It is to be noted that since both vectors $(q-v_o)$ and $(v_o-x)$ are in the same direction,  $\forall q\in\partial_{p+2} \mu(x)$, there exists $l\in\left[0,\rs{}-R_{v_o}\right]$ such that $q=v_o+l\left(v_o-x\right)/R_{v_o}$. On the other hand, since $\left(v_o-x\right)/R_{v_o}$ is a unit vector normal to the perimeter of the sensing disk (denoted by $n(v_o)$), the integral in (\ref{eq:gradient_obs}) is formulated as:
 \begin{align}
\nabla_{x}^{v_o}F(x)=
\left[
    \begin{matrix}
    0 & -1 \\
    1 & 0 
    \end{matrix}
\right]\frac{1}{R_{v_o}}n(v_o)
\int_{0}^{\rs{}-R_{v_o}}\varphi(q) ldl.
\end{align}
Using an approach similar to that in (\ref{eq:gradient_disk}), the above integral can be computed as a discrete summation over the $N_{p+2}$ points on $\partial_{p+2} \mu(x)$. Thus,
\begin{align}
\nabla_{x}^{v_o}F(x)=
\left[
    \begin{matrix}
    0 & -1 \\
    1 & 0 
    \end{matrix}
\right]\frac{\left(\rs{}-R_{v_o}\right)^2}{N_{p+2}R_{v_o}}n(v_o) \sum_{k=1\atop q_{k}\in\Pi'}^{N_{p+2}} k\varphi(q_k),
\end{align}
where $q_k=v_o+k\frac{\rs{}-R_{v_o}}{N_{p+2}}n(v_o)$, $k\in\mathbb{N}_{N_{p+2}}$.

In general, there may be more than one boundary segment generated, depending on the relative position of the sensor and the obstacles (characterized by vertices $v_o^{p+2},v_o^{p+3},\dots, v_o^{M}$). Consequently, the gradient of the objective function w.r.t. $x$ can be expressed as:
\begin{equation}
    \nabla_{x}F(x)=\nabla_{x}^D F(x)+\sum\limits_{k=p+2}^{M}\nabla_{x}^{v_o^{k}} F(x).
\end{equation}
With the above formulation, the high computational complexity of the line integral along the boundaries is replaced by a set of simple numerical summations affordable for sensors with limited computational capability.

After finding the gradient of the local weighted coverage and the decent direction $p_k$, the next step is to find the optimal moving step size such that the new candidate position provides the optimal value for $F(x)$. This is a line search problem that can be formulated as follows:
\begin{equation}
\alpha_{k}=\arg\min_{\alpha}F(x_{k}+\alpha p_{k}).\end{equation}
However, this may result in a point that is outside the region $\Pi$ or a point that the sensor cannot move to through a direct route due to the existence of some obstacles on its way. In such cases, the candidate point must be projected to the region $\Pi\cap\Phi(\xs{})$. Due to the similarity of this problem to the one studied in \cite{habibi2017}, the scaled gradient projection algorithm is used here. However, the non-convexity of the region caused by the heterogeneity of the network, the existence of static sensors, and the presence of obstacles, makes the problem more complex. 

\begin{thm}
Consider the sensor network $\Ss$ described in Section~\ref{sec2}, and let $\rc{i}\geq2\rs{i}, \forall i\in\mathbb{N}_n$. Then, the overall weighted coverage under the MMA strategy increases in each round until it reaches the steady state.
\end{thm}
\begin{Proof}
Although the CAMW-Voronoi diagram does not necessarily partition the field (due to the possible overlap between the regions), Theorem~\ref{thm: rc2rs} guarantees the distinctiveness of local covered areas. The proof follows now similarly to that of Theorem 2 in \cite{habibi2017}. 
\end{Proof}
It is worth mentioning that aside from the task of constructing the Voronoi region (which is common among all Voronoi-based approaches), the main computational complexity of the distributed MMA deployment strategy concerns deriving the local weighted coverage in each iteration. By using Delaunay triangulation of a region with $N$ points, the order of this complexity is $\mathcal{O}(N\log N)$ \cite{carstairs}. On the other hand, the number of points $N$, as a design parameter, determines the accuracy of integration, and hence, introduces a trade-off between the computation time and the precision of weighted coverage estimation. 

\section{Simulation Results}
\label{secex}
In this section, the performance of the MMA strategy is investigated in different scenarios. Due to the complexity and nonlinearity of the maximum weighted coverage problem, there is no tractable method to determine the globally optimal sensor configuration. Hence, the performance of the MMA strategy and two other methods is evaluated and compared using Monte Carlo simulations~\cite{wang2006,kwok2011}. Unlike the MMA strategy, the other methods are only applicable when the sensing field has a uniform priority function and no obstacles. Therefore, such a comparison is done only in Example 1. In all of the following examples, the communication radius of each sensor is assumed to be four times its sensing radius, which satisfies the condition of Theorem~\ref{thm: rc2rs}.

\textbf{Example 1. }
\label{ex1}
The goal of this example is to compare the performance of the MMA strategy with some other Voronoi-based approaches. In order for the scenario to provide credible information, the network and environment specifications have been chosen so that they mostly fit the constraints of other methods. For any test set, the sensors have been randomly deployed in a square field of $30\times30$m. One-third of the sensors are static and the remaining are mobile. The sensing radius is set to 2m for all static sensors and it is a random value between 2m and 4m for the mobile ones. The sensing radius of the static sensors is set to be less than or equal to that of the mobile sensors to reflect a real-world scenario in which sensors with insufficient power become permanently immobile at their last positions to minimize energy depletion.
Also, the field is assumed to have no obstacles, and a uniform priority function (i.e., $\varphi(q)\equiv 1$). Additionally, $\epsilon$ is set to $0.1 \text{m}^2$, meaning that the network reaches the steady state when no mobile sensor is able to improve its local coverage by at least $0.1 \text{m}^2$.
For Monte Carlo simulation, sensor networks of $12$, $21$, $30$, and $39$ sensors have each been tested for $20$ initial random configurations under the MMA, PCVF \cite{mahboubi2017}, and Minimax \cite{wang2006} methods for comparison. 

\begin{figure*}
\centering
     \begin{subfigure}{.24\textwidth}
         \includegraphics[trim=0 .7cm .7cm 0.4cm, clip, width=\textwidth]{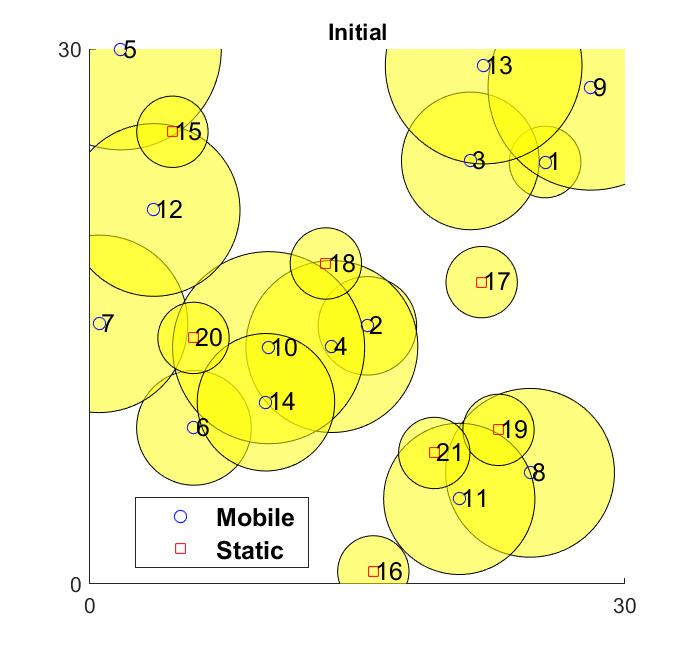}
     \end{subfigure}
     \begin{subfigure}{.24\textwidth}
         \includegraphics[trim=0 .7cm .7cm 0.4cm, clip, width=\textwidth]{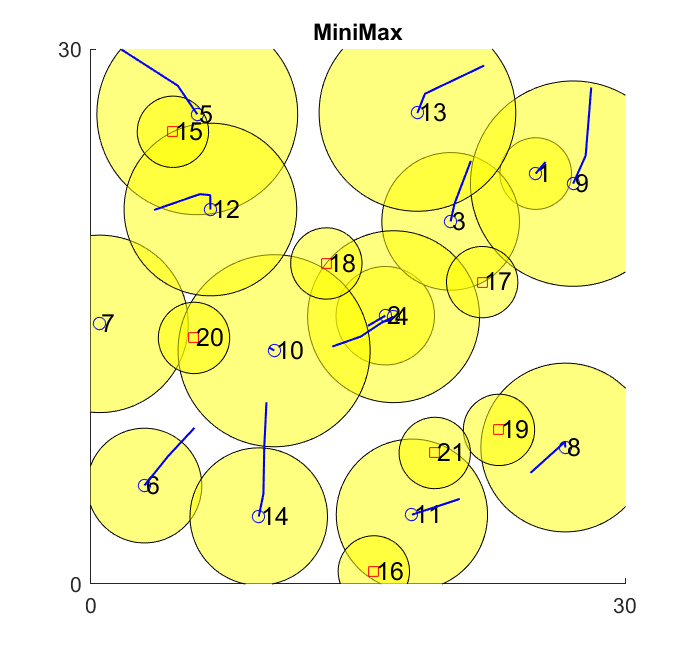}
     \end{subfigure}
     \begin{subfigure}{.24\textwidth}
         \includegraphics[trim=0 .7cm .7cm 0.4cm, clip,width=\textwidth]{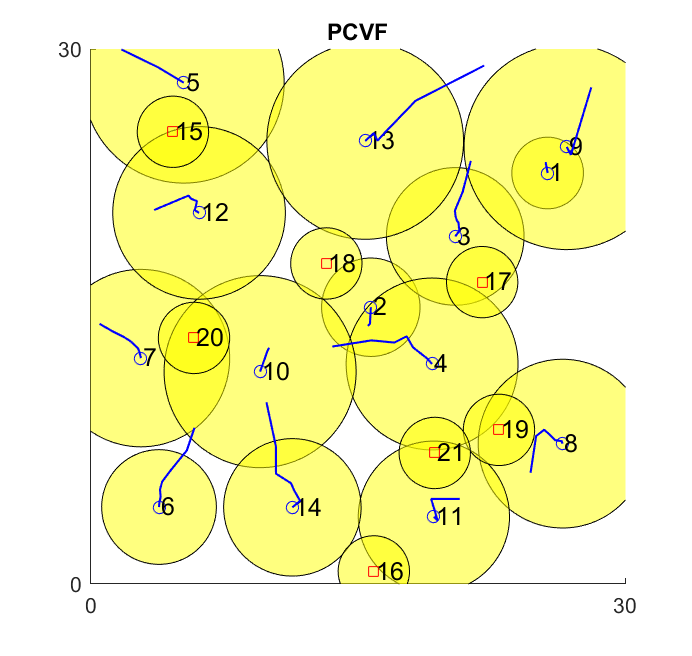}
     \end{subfigure}
     \begin{subfigure}{.24\textwidth}
         \includegraphics[trim=0 .7cm .7cm 0.4cm, clip, width=\textwidth]{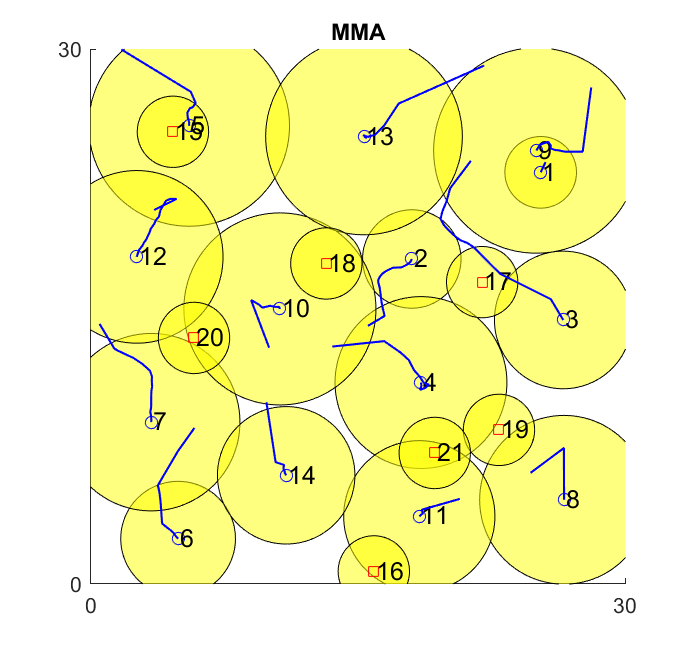}
     \end{subfigure}
     
        \caption{The initial and final configurations of the WSN under different strategies}
        \label{fig:sample_all}
\end{figure*}

One of the test results is shown in Fig.~\ref{fig:sample_all}, demonstrating that under the other two strategies, the network cannot avoid the unnecessary overlapping between the sensing ranges of mobile and static sensors since they do not take the static sensors into consideration initially for constructing regions.

\begin{figure}
    \centering
\includegraphics[trim=0.7cm .7cm 1.1cm 0.7cm, clip, width=.8\columnwidth]{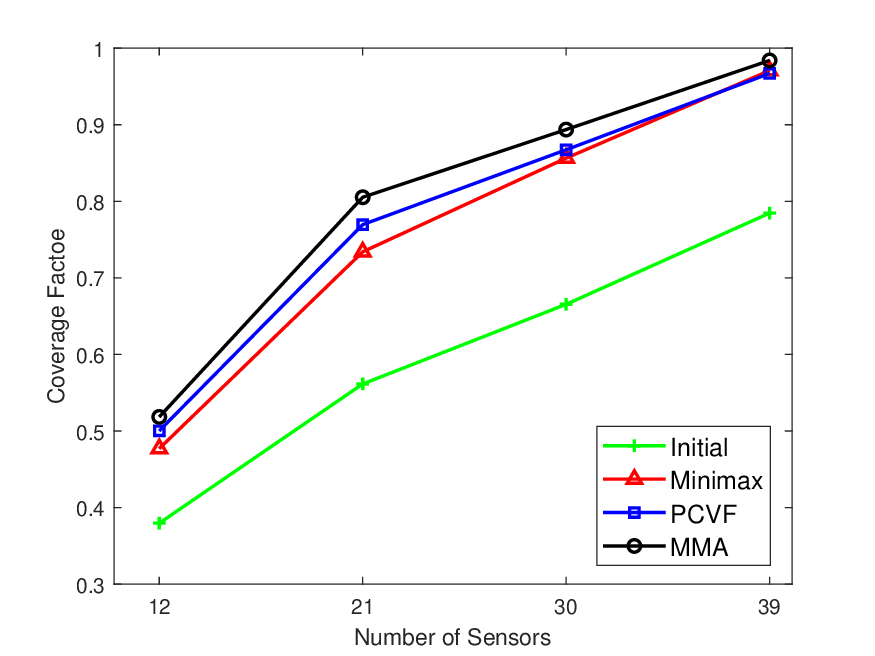}
    \caption{Final coverage factor for different numbers of sensors under different strategies}
    \label{fig:ex1_cov}
\end{figure}

\begin{figure}
    \centering
\includegraphics[trim=0.7cm .7cm 1.1cm 0.7cm, clip, width=.8\columnwidth]{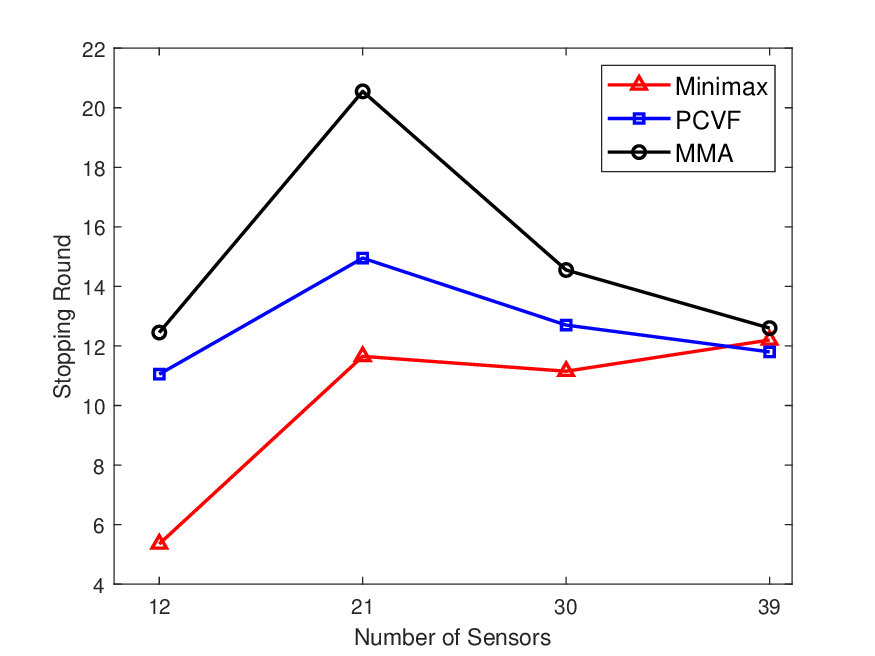}
\caption{Number of iterations before meeting the termination condition for different numbers of sensors under different strategies}
    \label{fig:ex1_stop}
\end{figure}

\begin{figure}
    \centering
\includegraphics[trim=0.7cm .7cm 1.1cm 0.7cm, clip,width=.8\columnwidth]{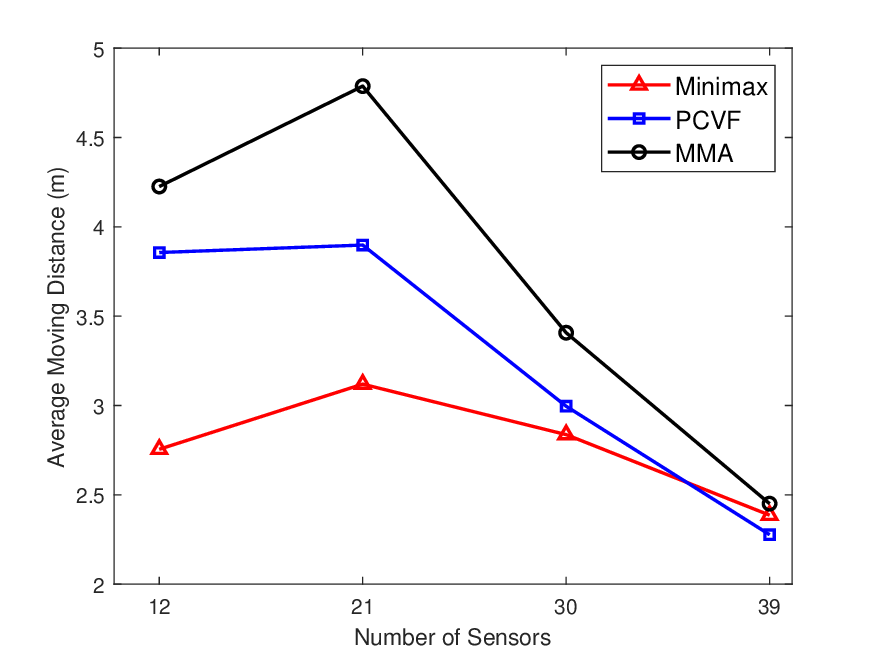}
    \caption{Average moving distance for different numbers of sensors under different strategies}
    \label{fig:ex1_dist}
\end{figure}

To assess the performance of the three methods, the average coverage factor, defined as the ratio of the weighted coverage of the sensor network to the weighted area of the field
versus the number of sensors in initial and final configurations are shown in Fig.~\ref{fig:ex1_cov}. This figure shows that the final coverage factor is always higher when applying the MMA method. On the other hand, Fig.~\ref{fig:ex1_stop} shows that the higher final coverage factor comes at the cost of a higher number of required iterations to reach the termination point. Also, based on Fig.~\ref{fig:ex1_dist}, the average moving distance for the sensors is also higher under the MMA strategy, meaning that each sensor travels a longer distance to find the appropriate location. 


Unlike the coverage factor which monotonously increases by increasing the number of sensors, the stopping round and average moving distance under the MMA strategy for networks with $21$ sensors are higher than those with smaller and larger networks. If the number of sensors is neither too low nor too high, the sensors move around constantly, searching for a configuration that results in the least overlapping and the most coverage.
In networks with a small number of sensors, movements may not be necessary as they may not affect coverage due to the low sensor density. In networks with a large number of sensors, also, extensive movements may not be necessary due to the potentially high overlap in sensors' covered areas.

\begin{figure}
    \centering
\includegraphics[trim=0.7cm .7cm 1.1cm 0.7cm, clip, width=.8\columnwidth]{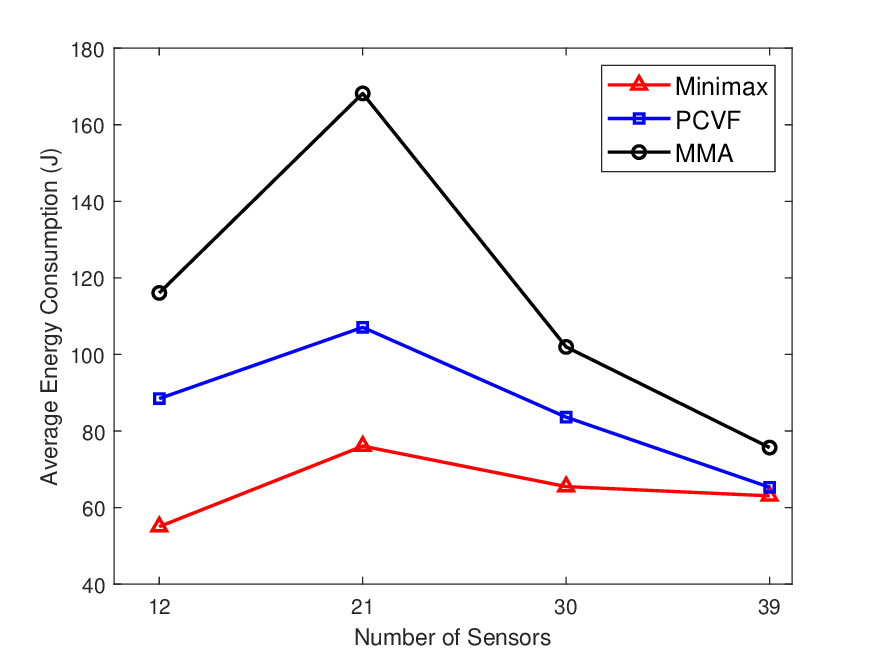}
    \caption{Average energy consumption for different numbers of sensors under different strategies}
    \label{fig:ex1_e}
\end{figure}

Energy consumption is another important factor that needs to be considered in evaluating the performance of sensor movement strategies because in real-world applications, sensors have a limited power supply. Let the energy that a sensor consumes to travel 1m (without
stopping) be $8.268$J \cite{yoon}. Also, suppose that the
amount of energy required to stop a sensor and then overcome
the static friction following a complete stop is equal to that required to travel $1$m and
$4$m, respectively \cite{wang2006}. The energy consumption for communication and sensing is assumed to be negligible compared to that for movement. Fig.~\ref{fig:ex1_e} gives the average energy consumed by each mobile sensor for different numbers of sensors.

Unlike the Minimax and PCVF strategies that act solely based on the relative positions of sensors in a non-prioritized field, the MMA method takes the area coverage of each sensor over its designated region into account. As a result, its improved final coverage factor comes at the cost of higher energy consumption and longer convergence time.

\begin{figure}
    \centering
    \includegraphics[trim=.8cm .5cm 0cm 1cm,width=\columnwidth]{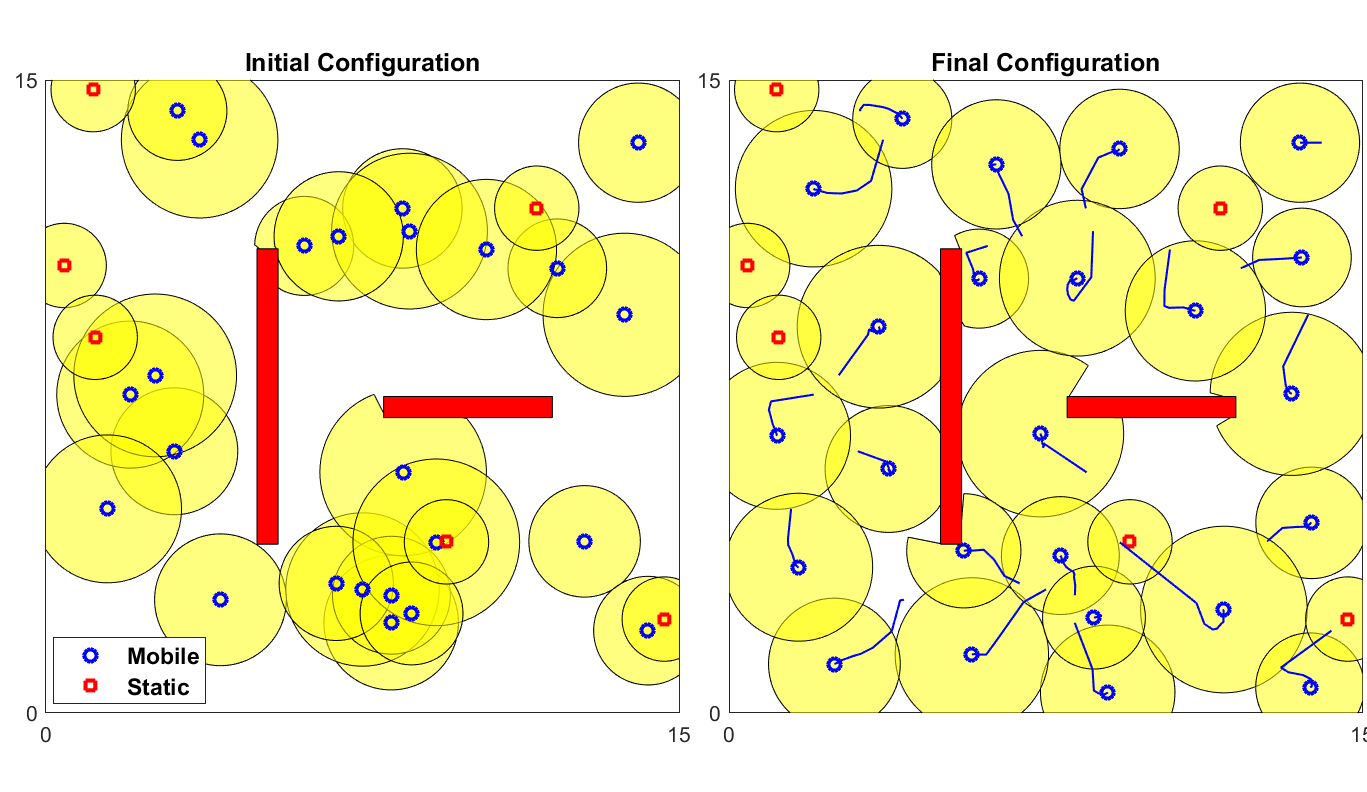}
    \caption{Performance of the MMA strategy for a network with mobile and static sensors in a  field with obstacles}
    \label{fig: ex3}
\end{figure}

\textbf{Example 2. }
In this example, the objective is to measure the performance of the MMA strategy in the presence of both static sensors and obstacles which is the main difference between this method and its predecessors. Consider 24 mobile sensors with random sensing radii between 1m and 2m, and 6 static sensors with sensing radii equal to 1m, randomly deployed in a square field of $15\times15$m with some obstacles as shown in Fig. ~\ref{fig: ex3}.
Similar to Example 1, the uniform priority function is considered.
Experiments are repeated $20$ times for different random initial configurations with and without obstacles. The initial and final configurations of the sensors for one of these tests are shown in Fig.~\ref{fig: ex3}. While the results demonstrate that the coverage increases in the final configuration,
as expected, they also show the negative impact of obstacles in the coverage performance, as evident from the average final coverage factor from 86\% to 69\%.

\begin{figure}
    \centering
\includegraphics[trim=1cm 0 0 0,clip,  width=.9\columnwidth]{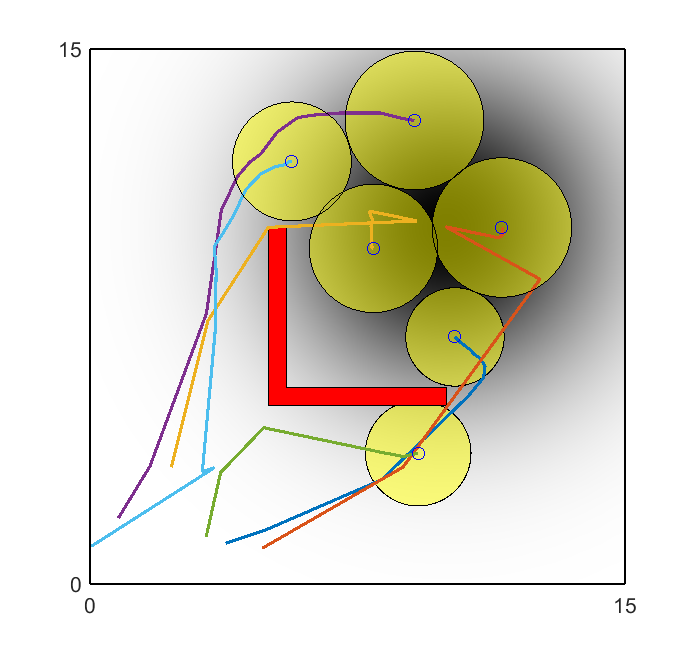}
    \caption{Performance of MMA strategy in a prioritized field with obstacles}
    \label{fig: ex2}
\end{figure}

\textbf{Example 3. }
In the last example, it is desired to demonstrate the performance of the MMA strategy in the presence of obstacles and a non-uniform priority function. The ROI is a square field of $15\times15$m with an obstacle as shown in Fig. ~\ref{fig: ex2}. The following priority function is used:
\begin{equation}
\label{eq:phi}
\varphi(q)=e^{-\alpha\left[(q_{1}-10)^{2}+(q_{2}-10)^{2}\right]},
\end{equation}
where $\alpha=0.02$. The above function has a peak value at the point $q=[10,10]^T$, and exponentially decays as moving farther from it. In Fig.~\ref{fig: ex2}, the darker spots indicate more important points to cover, according to the priority function. The sensor network includes 6 mobile sensors with sensing radii of $1.3$, $1.5$, $1.7$, $1.8$, $1.9$, and $2$ meters. The parameter $\epsilon$ is set to $0.05 \text{m}^2$ to enable the network to follow the priority function in areas where it has very low values. The sensors are initially deployed in the lower left corner of the field, meaning that their direct route toward the focal point of the priority function is blocked by the obstacle. 

Fig. ~\ref{fig: ex2} depicts the trajectories of the sensors and their final position, showing that they eventually cover the most important points by going around the obstacle. Furthermore, it shows that the sensors have taken the nearest routes by passing alongside the edge of the obstacle which can reduce the excessive traveling distance and energy consumption.

It is to be noted that multiple focal points can be modeled by a priority function equal to the sum of exponentials similar to (\ref{eq:phi}). A priority function with a greater $\alpha$ represents a faster decay. For a sensor network in such environments, it may be hard to detect the priority function and follow its gradient. Thus, the sensitivity of the network must be strengthened in such cases by decreasing the parameter $\epsilon$.

\section{Conclusion}
\label{secconc}
An iterative deployment strategy is proposed to maximize the weighted coverage of a network of mobile and static sensors with nonidentical sensing and communication radii over a field with obstacles. The objective is to develop a distributed approach, tasking every sensor to maximize the local coverage over the corresponding Voronoi region based on its interpretation of the environment and the information it obtains from its neighbors. The proposed method exploits a gradient-based approach to compute the optimal direction for every mobile sensor to move, considering the coverage priority of the points in the field, the static sensors' covered area, and the obstacles. The coverage efficacy of the method compared to the alternative approaches is demonstrated by simulations.
As future research direction, one can consider more practical settings, e.g., using more realistic sensing models (instead of a perfect disk), increased reliability using sweeping coverage or k-coverage schemes, and utilizing machine learning-based techniques or evolutionary algorithms for improved deployment performance.
{Future research directions involve potential adjustments to the problem, making the method applicable to scenarios involving practical constraints. This opens up the possibility of extending the findings to other WSN applications like distributed target tracking. Furthermore, the utilization of machine learning-based techniques and evolutionary algorithms can enhance deployment strategies for achieving the global optimum solution.}

\end{document}